# Electron transport in amorphous materials: from localization to predictive transport modeling

Yeonghun Lee[1,2,3,*]

[1] Department of Electronics Engineering, Incheon National University, Incheon 22012, Republic of Korea

[2] Department of Intelligent Semiconductor Engineering, Incheon National University, Incheon 22012, Republic of Korea

[3] Research Institute for Engineering and Technology, Incheon National University, Incheon 22012, Republic of Korea

**Corresponding author**

[*] Email: y.lee@inu.ac.kr

## Abstract

Amorphous materials are disordered solids without long-range structural order, making them useful systems for studying electron transport beyond the crystalline picture. This review discusses how structural disorder changes the spatial character of electron wavefunctions and how these changes govern carrier transport in amorphous materials, especially amorphous semiconductors. Basic concepts of Anderson localization, mobility edges, diffusive transport, and hopping transport are first reviewed, followed by representative amorphous semiconductors, including amorphous silicon and amorphous oxide semiconductors. Computational approaches are then discussed, from conventional Boltzmann and Green's-function-based transport theories to real-space Kubo–Greenwood simulations combined with molecular dynamics. The discussion focuses on localization, spectral broadening, finite-temperature lattice fluctuations, and electron–phonon interactions. Recent progress and remaining issues in predictive transport modeling of amorphous materials are then outlined.

## 1 Introduction

The relationship between structural disorder and electron transport has long been a major problem in condensed matter physics [1,2]. In crystalline solids, electron transport is commonly described within band theory, where charge carriers propagate as extended Bloch states and conductivity is largely determined by scattering processes. Anderson's work in 1958 showed that sufficiently strong disorder can localize wavefunctions through quantum interference [3]. This result demonstrated that disorder can determine the spatial character of wavefunctions, rather than simply acting as a source of scattering. Subsequent developments, including mobility edges, localization–delocalization transitions, and scaling theories of localization, further showed that transport in disordered systems is governed not only by scattering rates but also by the degree of wavefunction localization [4–7]. Extended states can support diffusive transport, whereas localized states give rise to hopping-type conduction [1,8,9]. Understanding how disorder controls localization and transport remains a key issue in condensed matter physics.

Amorphous materials encompass a broad range of systems, including semiconductors, oxides, metallic glasses, and chalcogenide materials [8,10,11]. Among these systems, amorphous semiconductors provide an important platform for examining the connection between atomic disorder, localization, and carrier transport [8,10,12,13]. Unlike crystalline solids, amorphous materials lack long-range periodic order while often retaining short-range chemical bonding. This structural character produces band-tail states and mobility edges that influence transport near the band edges [8,12,13]. Since the pioneering studies of amorphous silicon (a-Si) in the 1970s and 1980s, amorphous semiconductors have served as prototypical systems for investigating transport and optical processes in disordered solids [10,13,14]. More recently, amorphous oxide semiconductors (AOSs) have attracted considerable attention because relatively high electron mobility can be maintained even without long-range crystalline order [15–18]. These materials have renewed interest in the microscopic origin of transport in amorphous systems and have emphasized the need for theoretical and computational methods that can treat realistic structural disorder [16–18].

Theoretical and computational methods for describing transport in disordered materials have evolved in parallel with these experimental developments. Early studies relied mainly on analytical descriptions, whereas recent work increasingly employs atomistic simulations based on first-principles electronic-structure calculations [9,19–23]. Among the transport formalisms applicable to systems without translational symmetry, the Kubo–Greenwood (KG) formula [24–26] has played an important role because it expresses the electrical conductivity in terms of quantum-mechanical transitions between electronic states without requiring crystal momentum as a good quantum number [19,20]. Kubo–Greenwood calculations can incorporate structural disorder and quasi-static electron–phonon effects when combined with realistic amorphous structures and molecular-dynamics sampling [19,20]. Recent first-principles studies have begun to connect atomic structure and wavefunction localization directly to carrier mobility in amorphous materials [21–23]. In particular, recent work on amorphous oxide semiconductors has shown that atomistic modeling can capture band-like transport characteristics even in the absence of long-range crystalline order [23].

This review discusses disorder-induced localization and recent progress in atomistic transport modeling of amorphous materials. Section 2 summarizes the basic concepts of localization and transport in disordered solids. Section 3 discusses how these concepts appear in representative amorphous semiconductors, focusing on a-Si and AOSs. Section 4 reviews computational approaches for electron transport, from conventional Boltzmann and Green's-function-based transport theories to real-space KG simulations with molecular dynamics. Although representative examples are drawn from amorphous semiconductors, the transport formalisms discussed here are broadly relevant to amorphous materials in which structural disorder, localization, and thermal lattice fluctuations affect carrier transport.

## 2 Localization and transport in disordered solids

### 2.1 Anderson localization

The conventional description of electron transport in solids is based on band theory, where electrons propagate as extended Bloch states through a periodic crystal lattice. Within semiclassical transport theory, transport properties are then mainly determined by scattering processes. In this framework, disorder is often treated as a perturbation that modifies carrier lifetimes while leaving the extended nature of electronic states largely intact. A different perspective emerged from Anderson's seminal work, which demonstrated that sufficiently strong disorder can alter the character of the electronic states themselves [3]. To describe the effect of disorder on electronic states, Anderson introduced a tight-binding Hamiltonian with random on-site energies,

$$H = \sum_i \varepsilon_i c_i^\dagger c_i - t \sum_{\langle i,j \rangle} \left( c_i^\dagger c_j + \mathrm{h.c.} \right),$$

where $c_i^\dagger$ and $c_i$ are the electron creation and annihilation operators at site $i$, respectively, $\varepsilon_i$ is the random on-site energy, $t$ is the nearest-neighbor hopping integral, $i$ and $j$ label lattice sites, $\langle i,j \rangle$ denotes a pair of neighboring sites, and h.c. denotes the Hermitian-conjugate term. In this model, disorder is represented by random site energies while the hopping integral is taken to be uniform. In the absence of disorder, the eigenstates of this Hamiltonian are extended throughout the crystal and can be described by Bloch wavefunctions. As the disorder strength increases, however, electrons undergo multiple scattering from the random potential landscape. The resulting quantum interference between different propagation paths can suppress diffusion and eventually localize the wavefunctions, showing that disorder can govern the nature of electronic states, not only the scattering rate [1,3].

A consequence of Anderson localization is the emergence of spatially localized states whose amplitudes decay exponentially away from a localization center [1]. Such localized states are characterized by a localization length $\xi$, which quantifies the spatial extent of the wavefunction through

$$|\psi(\mathbf{r})| \sim \exp\left(-\frac{|\mathbf{r}-\mathbf{r}_0|}{\xi}\right),$$

where $\psi(\mathbf{r})$ is the wavefunction amplitude at the position vector $\mathbf{r}$, $\mathbf{r}_0$ is the localization center, and $\xi$ is the localization length. Unlike extended states, localized states cannot support diffusive conduction because electrons remain confined within finite regions of space. Consequently, the transport properties of a disordered material depend not only on the scattering rate but also on whether the relevant wavefunctions are extended or localized. These two types of states may coexist within the same energy band; states near the band center often remain extended, whereas those near the band edges become localized due to disorder. The energy separating extended and localized states is known as the mobility edge [5]. This mobility-edge picture is schematically illustrated in Fig. 1. When the Fermi level lies within the extended-state region, electrons can propagate through the system and conventional diffusive transport is possible. In contrast, when the Fermi level falls within the localized-state region, electronic conduction is suppressed and transport must proceed through alternative mechanisms involving transitions between localized states. As the disorder strength increases, localized states occupy a larger fraction of the energy spectrum, causing the mobility edge to move toward the band center. In sufficiently disordered systems, all states at the Fermi energy may become localized, resulting in the disappearance of metallic conduction and the onset of insulating behavior. This disorder-driven metal–insulator transition highlights the central role of the spatial character of electronic states in transport [4,6,7].

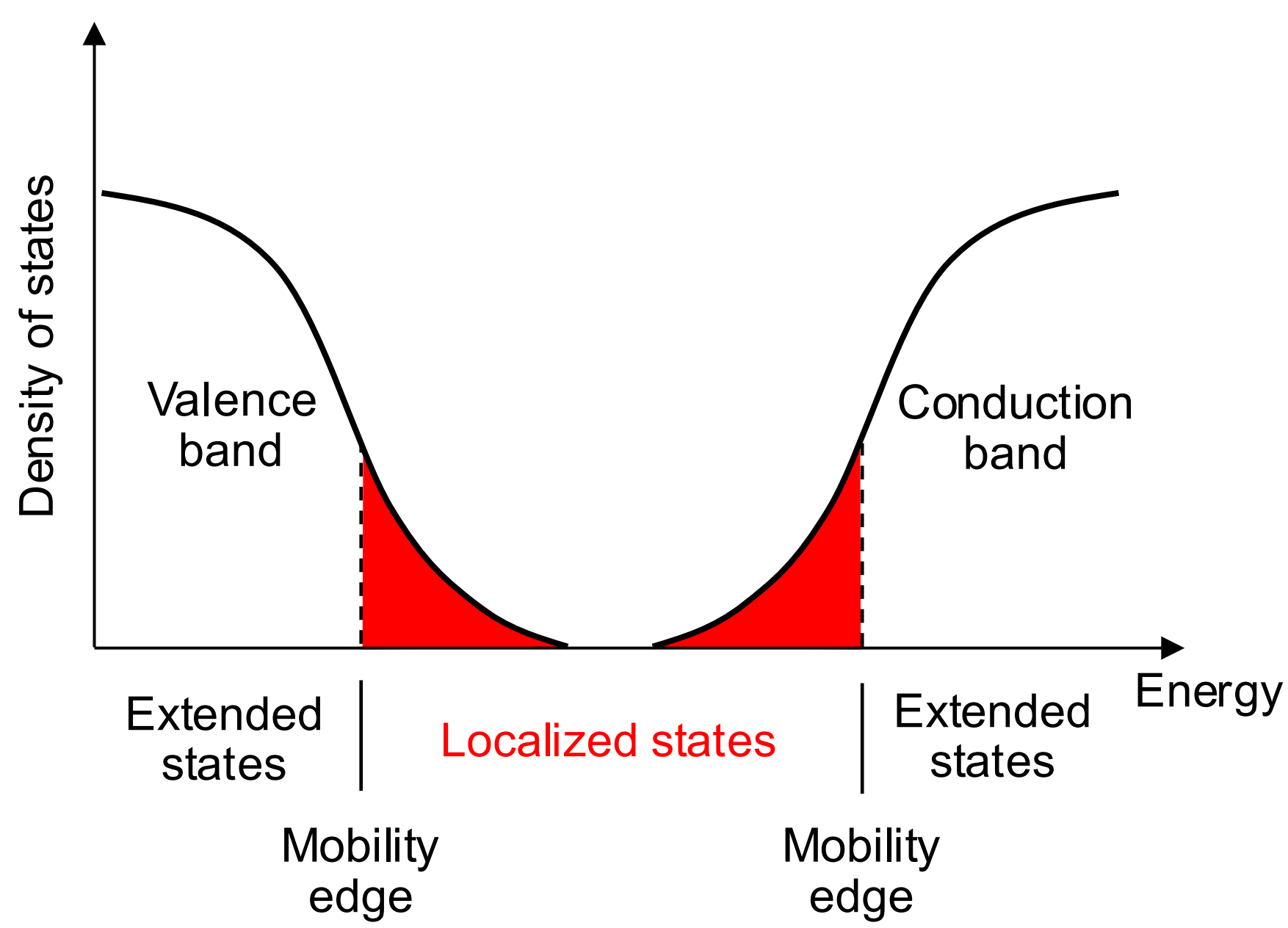

Fig. 1 Schematic illustration of electronic states in a disordered solid. Structural disorder produces localized states near the band edges, whereas extended states remain inside the bands. The mobility edge separates localized and extended states.

**2.2 Transport mechanisms in disordered solids**

As discussed in the previous section, disorder can give rise to both extended and localized states [1,6]. Electron transport in disordered solids can be classified into two limiting regimes:

diffusive transport through extended states and hopping transport between localized states. In the diffusive regime, transport occurs through extended states on length scales larger than the mean free path. Repeated scattering by disorder, phonons, or other electrons then leads to the diffusive propagation of electron wave packets, resulting in a finite diffusion coefficient and finite conductivity. Although disorder reduces the carrier mean free path, extended states can still support diffusive transport. Consequently, the conductivity in the diffusive regime is governed primarily by scattering processes rather than by wavefunction localization [1].

The Ioffe–Regel criterion provides a useful estimate of the limit of the semiclassical quasiparticle description of such transport: when the carrier mean free path becomes comparable to the electron wavelength or to the interatomic spacing, the quasiparticle-based transport picture loses its validity [27]. In crystalline systems, this condition is often expressed as $kl{\sim}1$, where $k$ is the magnitude of the wave vector and $l$ is the mean free path. In amorphous materials, however, crystal momentum is not a well-defined quantum number, so this criterion should be understood as a qualitative indicator of the breakdown of the semiclassical quasiparticle description rather than as a strict momentum-space condition.

In contrast, hopping transport arises when the relevant electronic states are spatially localized. Because localized states do not extend throughout the system, charge carriers cannot propagate through the material via conventional band-like transport. Instead, transport occurs through a sequence of transitions between localized states. These transitions are typically assisted by phonons, which supply the energy needed to bridge the energy mismatch between localized states and enable carriers to hop between localized states at different positions. As a result, hopping transport differs from conventional diffusive transport through extended states and is typically characterized by thermally activated conductivity [8,9]. Depending on the distribution of localized states and the temperature range, different hopping regimes may emerge. At relatively high temperatures, carriers often hop to nearby localized states because thermal energy can overcome the associated activation energy. At lower temperatures, variable-range hopping (VRH) can become dominant; in this regime, carriers hop to more distant states with smaller energy differences [8,9,28,29]. These hopping regimes are governed by the localization length, the density of localized states, and phonon-assisted transition probabilities, rather than by band velocities and momentum relaxation times. Thus, hopping conduction provides the transport channel when carriers occupy localized band-tail or defect states [8,9,29]. In practical materials, electron transport is often more complex than these idealized limiting cases because localized and extended states may coexist within the same energy spectrum, particularly near the mobility edge [5,6]. As a result, transport can involve both diffusive and hopping contributions, with the dominant mechanism depending on temperature, carrier concentration, and electronic states. This coexistence of extended and localized states is central to transport in amorphous semiconductors, where structural disorder produces band-tail states near the band edges [8,10].

## 3 Electron transport in amorphous semiconductors

### 3.1 Electronic structure of amorphous semiconductors

Unlike crystalline semiconductors, amorphous semiconductors lack long-range periodic order while often retaining local bonding environments that resemble those of the corresponding crystalline phases [8,10]. This combination of short-range order and long-range disorder is a defining structural feature of amorphous semiconductors. Variations in bond lengths, bond angles, and atomic coordination introduce spatial fluctuations in the local potential landscape, which modifies the band edges, gap states, and spatial character of wavefunctions [8,10,12]. As a result, the electronic structure of amorphous semiconductors should not be viewed simply as a broadened version of the crystalline electronic structure; it is reshaped by the underlying disordered atomic network.

One of the prominent consequences of this disorder is the formation of band-tail states [8,10,12,13]. In crystalline semiconductors, the valence- and conduction-band edges are well defined by extended Bloch states. In amorphous semiconductors, however, structural disorder broadens these band edges and produces localized tail states that extend into the band gap (see Fig. 1). Their energetic distribution and degree of localization influence the electronic properties of amorphous semiconductors, particularly near the band edges where charge carriers are often introduced by doping, gating, or optical excitation. In addition to band-tail states, amorphous semiconductors can contain defect states associated with coordination defects, vacancies, dangling bonds, floating bonds, or compositional disorder [8,10]. Such defect levels are typically located deeper in the band gap than band-tail states and can act as carrier traps or recombination centers. In relatively defect-free amorphous networks, states near the band edges are mainly band-tail states produced by structural disorder, whereas coordination defects or chemical disorder can introduce deeper gap states. Distinguishing between these two types of disorder-induced states is therefore important for interpreting transport and optical measurements in amorphous semiconductors.

For band-edge states, the mobility edge separates extended states inside the bands from localized tail states near the band edges [5,8,12]. Deep defect states within the gap are also localized, but they are usually treated separately as trap or recombination states. More broadly, amorphous semiconductors may be viewed as disorder-broadened energy spectra containing extended states, band-tail states, and defect states with different degrees of localization [8,10,12]. In modern atomistic descriptions, these electronic-structure features are understood as direct consequences of the underlying disordered atomic network [19]. This atomistic viewpoint provides the basis for comparing how different bonding characters lead to distinct localization and transport behaviors in a-Si and AOSs.

### 3.2 Amorphous silicon and oxide semiconductors

Among the various classes of amorphous semiconductors, a-Si and AOSs have attracted particular attention owing to their contrasting electronic structures and transport characteristics [8,10,15–18]. Both lack long-range crystalline order and exhibit disorder-induced electronic states, yet their carrier transport properties differ substantially. Amorphous silicon is the prototypical covalently bonded amorphous semiconductor and has long served as a model system for studying the effects of disorder on electronic properties [8,10,14]. Despite retaining the fourfold coordination characteristic of crystalline silicon on average, the amorphous network exhibits significant variations in bond lengths and bond angles, together with coordination defects such as dangling bonds [10]. Because the valence- and conduction-band states are formed from directional $sp^3$ orbitals, structural distortions strongly perturb these states and promote localization near the band edges [8,10]. Consequently, a-Si exhibits pronounced band-tail states near the band edges [8,10,12]. Charge transport is therefore strongly influenced by localization effects, particularly at low carrier concentrations and near the mobility edge [8,10,12–14].

In contrast, amorphous oxide semiconductors represent a distinct class of materials in which relatively high electron mobility can be maintained despite the absence of long-range order [15–18]. Typical examples include amorphous indium gallium zinc oxide (a-IGZO) and amorphous indium oxide (a-$In_2O_3$) [16–18]. Unlike a-Si, the conduction-band states in many AOSs are derived primarily from spatially extended metal $s$ orbitals. Because these orbitals are nearly isotropic, their overlap is less sensitive to bond-angle distortions than that of directional covalent orbitals [16–18]. As a result, conduction-band states often remain relatively delocalized even in highly disordered structures, leading to significantly higher electron mobilities than those observed in a-Si [15–18]. The extended nature and nearly isotropic overlap of metal $s$ orbitals are therefore widely regarded as key factors underlying the relatively high electron mobility of AOSs [16–18]. This conduction-state connectivity in oxide semiconductors is schematically illustrated in Fig. 2. However, this orbital-based picture may be incomplete. In particular, studies that emphasize the coupling between In $s$ and O $p$ orbitals suggest that carrier transport should also be discussed in terms of the metal–oxygen network and its sensitivity to structural disorder [30,31].

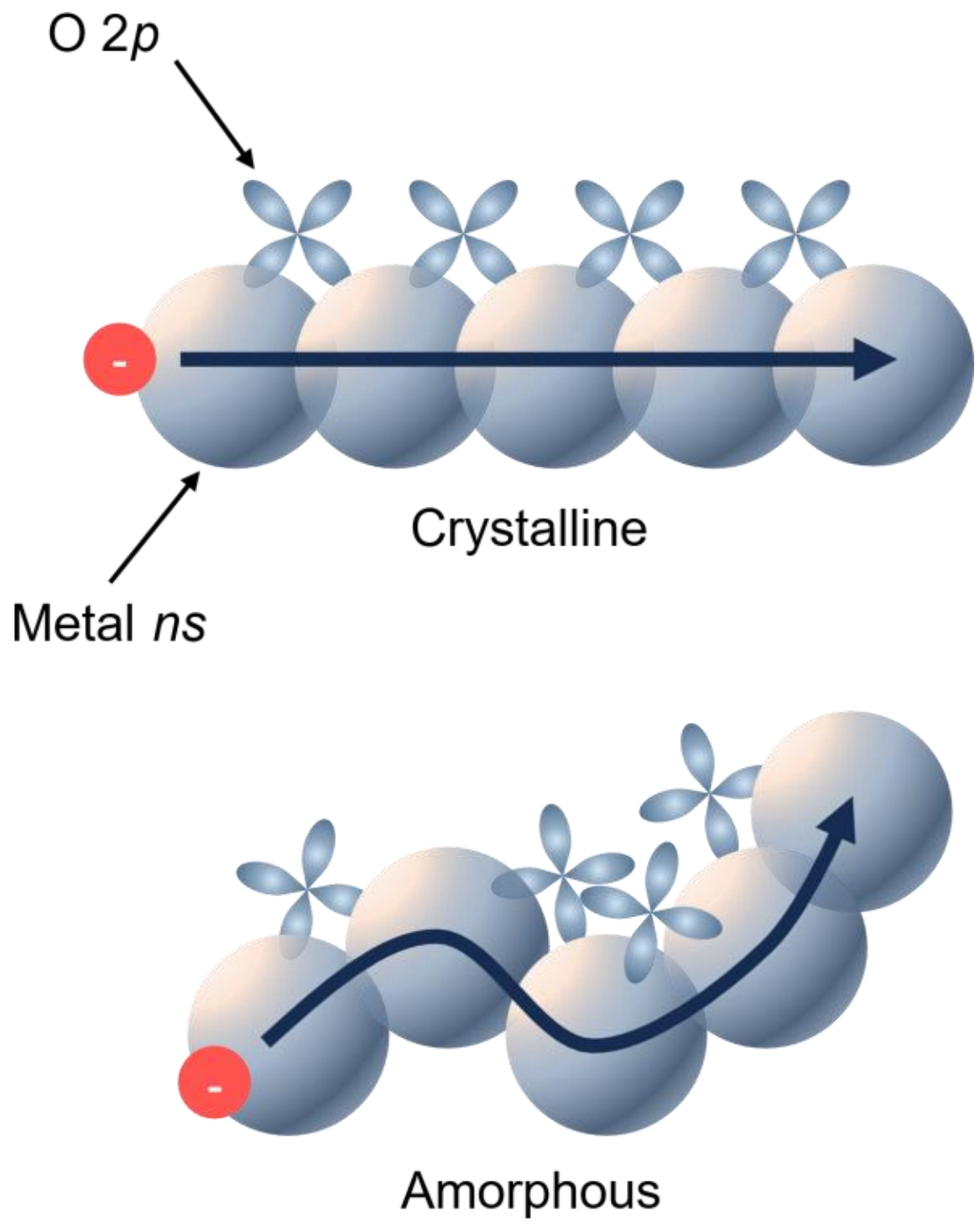

Fig. 2 Schematic comparison of conduction-state connectivity in crystalline and amorphous oxide semiconductors. In crystalline oxide semiconductors, periodic metal–oxygen networks support extended conduction states. In amorphous oxide semiconductors, long-range periodicity is lost, but conduction-state connectivity mediated by metal orbitals and the metal–oxygen network can remain preserved, allowing conduction states to retain relatively delocalized character despite structural disorder.

Even in high-mobility AOSs, the preservation of extended conduction states does not imply the absence of disorder effects. Structural disorder still produces band-tail states near the band edges, and transport can become sensitive to these states when the carrier energy approaches the mobility edge [16–18]. Experimental signatures such as Meyer–Neldel behavior and crossovers between band-like and hopping-like transport show that carrier mobility cannot be explained solely by the isotropic overlap of metal *s* orbitals [17]. Instead, it reflects the combined influence of extended conduction states, localized tail states, and finite-temperature lattice fluctuations. Recent atomistic studies further support this picture by linking mobility trends to local structural disorder and wavefunction localization [22,23]. Thus, AOSs should be viewed as disorder-tolerant, not disorder-free, amorphous semiconductors. The comparison between a-Si and AOSs shows that amorphous disorder affects transport through the orbital character and connectivity of the band-edge states.

Quantitative mobility calculations for AOSs, particularly a-$In_2O_3$, have employed several material-specific approaches because conventional Boltzmann transport theory cannot be directly applied to amorphous electronic states for which crystal momentum is not a well-defined quantum number. Lee *et al.* [21] combined first-principles calculations of amorphous structures with a mobility–localization relation established from model NEGF calculations, using the inverse participation ratio to estimate the localization length. Jang *et al.* [22] represented structural disorder as an effective scattering source by converting variations in Bader charges into a density of charged structural-disorder scattering centers, which was then included together with polar-optical-phonon and ionized-impurity scattering in an effective-mass Boltzmann transport model. More recently, Jankousky *et al.* [23] represented a-$In_2O_3$ as an ensemble of small periodic local environments and obtained an effective conduction-band dispersion and disorder-induced spectral linewidth from ensemble-averaged quasiparticle self-consistent $GW$ (QS$GW$) spectral functions. The effective mass and relaxation time were extracted from the band dispersion and spectral linewidth, respectively, and the disorder-limited mobility was then estimated using a simple Drude relation.

These approaches provide different routes to quantitative mobility modeling, but each involves specific approximations. The approach of Ref. [21] relies on a separation between disorder-induced localization and the scattering length scale, rather than treating both effects directly within the amorphous Hamiltonian. The approach of Ref. [22] assumes an effective-mass band and maps structural disorder onto impurity-like scattering, substantially simplifying the description of the complex amorphous structure. The approach of Ref. [23] provides a first-principles description of the effective band structure and disorder-induced spectral broadening, but the mobility is ultimately evaluated using a simple Drude relation. It also relies on representing the amorphous solid as an ensemble of local environments and does not explicitly include dynamical electron–phonon scattering. These studies illustrate both recent progress and the continuing need for transport approaches that can treat structural disorder, localization, and electron–phonon interactions within a unified framework.

## 4 Computational approaches for transport in amorphous materials

### 4.1 Conventional transport theories

Theoretical descriptions of electron transport have traditionally been developed for crystalline solids, where electronic states are characterized by well-defined band dispersions and crystal momentum. In such systems, transport properties are governed by the interplay between electronic structure and scattering processes, including electron–phonon interactions and defects. Two major formalisms have served as foundations of transport theory: semiclassical Boltzmann transport theory and Green's-function-based approaches [24,32–34]. These methods have successfully described carrier transport in many crystalline semiconductors and metals. However, their direct application to amorphous materials is limited because the absence of long-range translational symmetry makes crystal momentum and momentum-resolved scattering rates difficult to define, while structural disorder can also change the spatial character of the wavefunctions.

One widely used method is semiclassical Boltzmann transport theory. In this framework, electrons are treated as wave packets propagating through extended Bloch states labeled by crystal momentum $\mathbf{k}$ and band index $n$. Within the relaxation-time approximation, the dc conductivity tensor is written as [32]

$$\sigma_{\alpha\beta}^{\mathrm{dc}} = e^2 \sum_n \int \frac{d\mathbf{k}}{(2\pi)^3} v_{n\alpha}(\mathbf{k}) v_{n\beta}(\mathbf{k}) \tau_n(\mathbf{k}) \left(-\frac{\partial f(E)}{\partial E}\right).$$

Here, $e$ is the elementary charge, $n$ and $\mathbf{k}$ denote the band index and crystal-momentum vector, respectively, $v_{n\alpha}(\mathbf{k})$ is the group velocity along the Cartesian direction $\alpha$, $\tau_n(\mathbf{k})$ is the relaxation time, and $f(E)$ is the Fermi–Dirac distribution. The energy derivative of $f(E)$ is evaluated at $E = \varepsilon_n(\mathbf{k})$. The indices $\alpha$ and $\beta$ denote Cartesian components of the conductivity tensor. In modern first-principles calculations, scattering rates due to electron–phonon interactions can be computed explicitly, enabling quantitative predictions of carrier mobility and conductivity in crystalline materials [35–38]. Nevertheless, this approach relies on translational symmetry and well-defined crystal momentum. When disorder becomes strong and electronic states acquire localized character, these assumptions gradually break down.

Beyond semiclassical transport theory, a more general description can be obtained within linear-response theory [24,33,34]. In the Kubo formula, the electrical conductivity is expressed in terms of equilibrium current–current correlation functions rather than carrier trajectories and scattering events [24,34]. This formula provides a fully quantum-mechanical description of transport and forms the basis of Green's-function-based transport theories. Because disorder and interaction effects can be incorporated systematically through correlation functions and self-energies, linear-response theory has become an important theoretical tool for describing transport in condensed matter systems. In practical implementations, the Kubo formula is expressed using Green's functions, which provide a convenient framework for incorporating disorder and interaction effects through the self-energy [33,34]. The disorder-averaged retarded Green's function operator may be written as

$$\hat{G}^{\mathrm{R}}(E) = \left[(E + i0^{+})\hat{I} - \hat{H}_0 - \hat{\Sigma}^{\mathrm{R}}(E)\right]^{-1},$$

where $0^{+}$ is a positive infinitesimal that selects the retarded Green's function, $\hat{I}$ is the identity operator, $\hat{H}_0$ is the unperturbed electronic Hamiltonian, and $\hat{\Sigma}^{\mathrm{R}}(E)$ is the retarded self-energy arising from disorder. Depending on the strength of disorder, different approximations can be used to evaluate $\hat{\Sigma}^{\mathrm{R}}(E)$.

In the Born approximation (BA), disorder is treated perturbatively. For a disorder potential with zero configurational average, or after its average has been absorbed into the reference Hamiltonian, the leading nonvanishing self-energy is given by [33,34]

$$\hat{\Sigma}^{\mathrm{R}}_{\mathrm{BA}}(E) = \left\langle \hat{V}\hat{G}^{\mathrm{R}}_0(E)\hat{V} \right\rangle_{\mathrm{dis}},$$

where $\hat{V}$ is the disorder potential, $\hat{G}^{\mathrm{R}}_0(E)$ is the retarded Green's function of the reference system, and $\langle \dots \rangle_{\mathrm{dis}}$ denotes averaging over disorder configurations. The Born approximation provides a useful description of weak disorder and weak scattering, but its accuracy decreases as disorder becomes stronger. For stronger disorder, self-consistent effective-medium approaches are often employed. Among them, the coherent potential approximation (CPA) has become a widely used method for treating substitutional or configurational disorder [39,40]. In this approach, the disordered system is replaced by an effective medium characterized by a site-independent coherent potential (self-energy) $\hat{\Sigma}^{\mathrm{R}}_{\mathrm{CPA}}(E)$, so that all sites become equivalent in the effective medium. The coherent potential is determined self-consistently from the condition

$$\langle \hat{T}_i(E) \rangle_{\mathrm{dis}} = 0,$$

where $\hat{T}_i(E)$ is the single-site $T$-matrix obtained by replacing an effective-medium site with one of the possible local site potentials. The corresponding $T$-matrices are averaged over the possible site potentials according to their probabilities, and the CPA condition requires this disorder-averaged excess scattering to vanish. Compared with the BA, the CPA can describe disorder-induced spectral broadening in a nonperturbative but computationally tractable manner, providing an effective-medium description of how disorder modifies energy spectra and transport properties. In the weak-scattering limit, the CPA recovers the Born result to leading order in the disorder strength [40]. For stronger disorder, however, it provides a self-consistent effective-medium description beyond perturbative scattering. The BA and CPA treatments of disorder are schematically compared in Fig. 3(a). However, because the CPA is based on disorder-averaged quantities, it does not capture Anderson localization itself. Typical medium theory (TMT) addresses this limitation by introducing the geometrically averaged local density of states, known as the typical density of states, as a local order parameter for Anderson localization [41]. Unlike the conventional disorder-averaged density of states, the typical density of states vanishes as electronic states become localized.

Once the self-energy is obtained, the resulting states can be characterized through the spectral function projected onto the $n$-th eigenstate $|n\rangle$ [33,34],

$$A_n(E) = -\frac{1}{\pi} \mathrm{Im} \langle n | \hat{G}^{\mathrm{R}}(E) | n \rangle.$$

The spectral function describes how spectral weight is distributed in energy and how it is shifted and broadened by self-energy effects (see Fig. 3(b)). In weak-scattering or quasiparticle regimes, this broadening is often represented by a Lorentzian peak. In more general disordered systems, however, the spectral function can deviate from a Lorentzian form because the self-energy may depend strongly on energy. In quasiparticle regimes, the spectral width is associated with a finite lifetime and therefore provides a microscopic connection between disorder, scattering, and transport. Green's-function-based transport theories connect these spectral quantities to conductivity through linear-response expressions involving current or velocity correlation functions. These formulations provide a link between microscopic electronic structure and measurable transport properties.

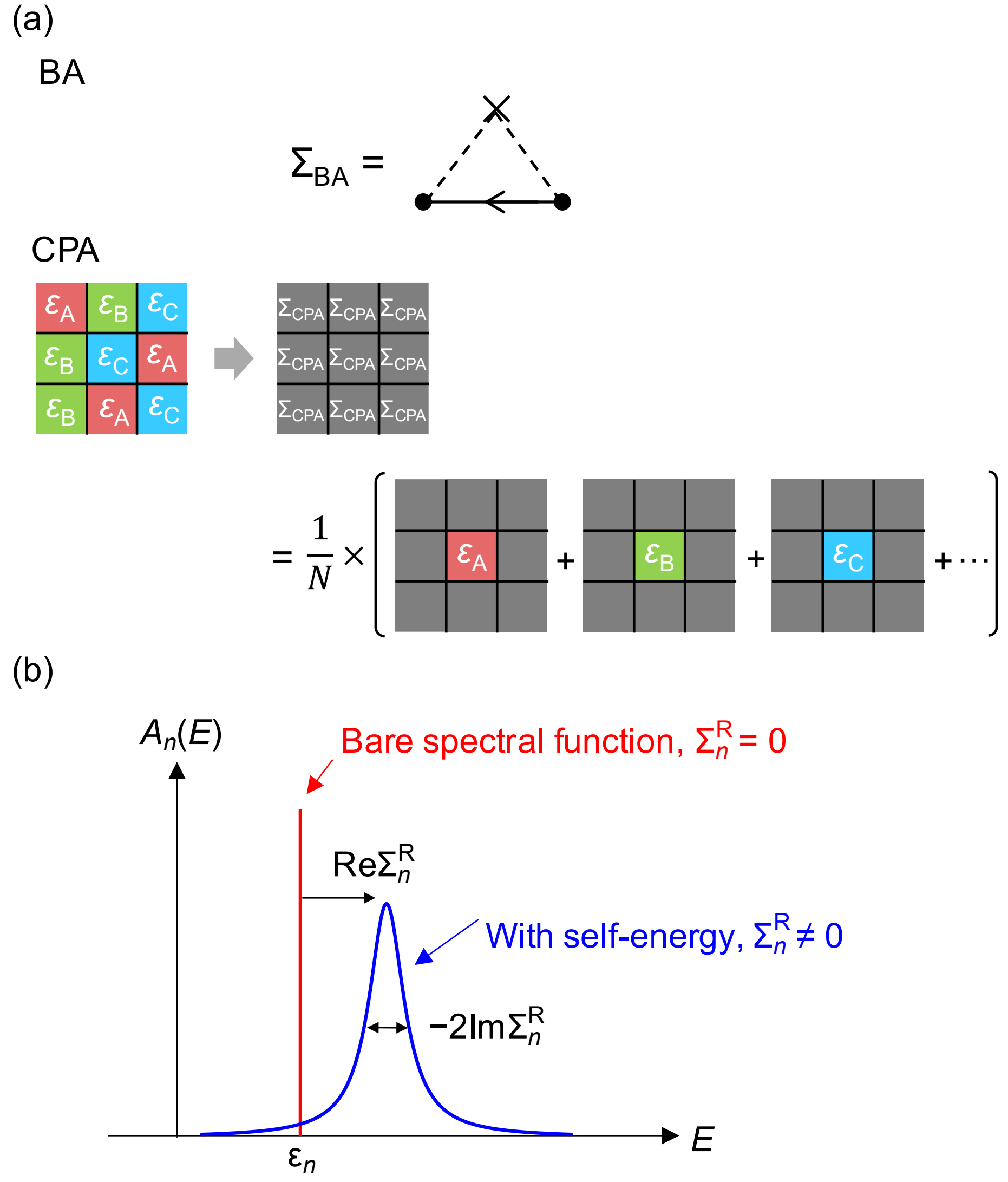

Fig. 3 (a) Schematic comparison of BA and CPA treatments of disorder. In the impurity-averaged BA diagram, the solid line represents the retarded Green's function of the reference system, and the filled circles represent scattering vertices associated with the disorder potential. The dashed lines connected to the cross represent averaging over the position of the same impurity, which restores translational invariance and hence momentum conservation. The diagram therefore represents an impurity-averaged double-scattering process that gives the BA self-energy $\Sigma_{BA}$ [33,34]. In the CPA panel, $\varepsilon_A$, $\varepsilon_B$, and $\varepsilon_C$ represent the possible local site potentials in the disordered lattice, whereas $\Sigma_{\mathrm{CPA}}$ represents the site-independent coherent potential (self-energy) of the effective medium. Thus, all sites are equivalent in the effective medium. The lower diagrams illustrate replacing one effective-medium site with each possible local potential. The corresponding single-site scattering matrices are then averaged over the possible site potentials according to their probabilities, and the coherent potential is determined by requiring $\langle \hat{T}_i(E) \rangle_{\mathrm{dis}} = 0$. (b) Schematic illustration of spectral broadening induced by the self-energy. The real part of the self-energy shifts the spectral peak, whereas the imaginary part determines its width and is associated with a finite lifetime. When the self-energy is approximately diagonal and weakly energy dependent over the spectral peak, the Lorentzian full width at half maximum (FWHM) is approximately given by $-2\mathrm{Im}\Sigma_n^{\mathrm{R}}(E) = -2\mathrm{Im}\langle n|\hat{\Sigma}^{\mathrm{R}}(E)|n\rangle$.

Despite their success, both Boltzmann transport theory and the perturbative or effective-medium Green's-function approaches discussed above face limitations when applied to amorphous materials [1,8]. The absence of long-range translational symmetry eliminates crystal momentum as a good quantum number, making band velocities and momentum-dependent scattering rates difficult to define. In addition, structural disorder can generate localized states and mobility edges, leading to transport mechanisms that cannot be described solely in terms of scattering between extended Bloch states [5–8]. Effective-medium approaches such as the CPA capture averaged disorder effects through self-energy corrections and spectral broadening, but they do not explicitly represent the microscopic atomic configurations of realistic amorphous structures [39,40,42]. These limitations motivate the use of transport methods based directly on atomistic structures, which are particularly important for amorphous materials. Among such approaches, real-space simulations based on the KG formula provide a direct route for evaluating transport without relying on crystal momentum [19,20].

### 4.2 Real-space Kubo–Greenwood transport simulations

In real-space implementations of the KG formula [19,20], the electrical conductivity is expressed in terms of transitions between eigenstates, without relying on well-defined Bloch wave packets. The conductivity is commonly formulated first at finite frequency, from which the dc conductivity can be obtained in the low-frequency limit. Within the independent-particle approximation, the real part of the ac conductivity tensor is written as [24–26]

$$\sigma_{\alpha\beta}(\omega) = \frac{2\pi e^2 \hbar}{\Omega} \int dE \frac{f(E) - f(E + \hbar\omega)}{\hbar\omega} \mathrm{Tr}\left[\hat{v}_\alpha \delta\left(E - \hat{H}\right) \hat{v}_\beta \delta\left(E + \hbar\omega - \hat{H}\right)\right],$$

where $\Omega$ is the system volume, $\hat{H}$ is the electronic Hamiltonian, $\hat{v}_\alpha$ and $\hat{v}_\beta$ are velocity operators along the Cartesian directions $\alpha$ and $\beta$, respectively, and $\omega$ is the angular frequency. The factor 2 accounts for spin degeneracy in nonmagnetic calculations. In the eigenstate representation, this expression becomes

$$\sigma_{\alpha\beta}(\omega) = \frac{2\pi e^2 \hbar}{\Omega} \sum_{n,m} \frac{f(\varepsilon_n) - f(\varepsilon_m)}{\varepsilon_m - \varepsilon_n} \langle n|\hat{v}_\alpha|m\rangle \left\langle m\middle|\hat{v}_\beta\middle|n\right\rangle \delta(\varepsilon_m - \varepsilon_n - \hbar\omega),$$

where $\varepsilon_n$ and $|n\rangle$ denote the eigenvalues and eigenstates of the disordered Hamiltonian, respectively, $\langle n|\hat{v}_\alpha|m\rangle$ and $\left\langle m\middle|\hat{v}_\beta\middle|n\right\rangle$ are velocity-operator matrix elements, and the Dirac delta function imposes energy conservation, $\varepsilon_m - \varepsilon_n = \hbar\omega$. This form is useful for amorphous materials because the conductivity is determined directly by the spatial character of the electronic states and the velocity matrix elements connecting them. Extended, weakly localized, and strongly localized states can therefore be treated within the same formalism. In amorphous-material simulations, the KG formula is usually evaluated using large supercells [19,20]. The Hamiltonian may be obtained from first-principles calculations or tight-binding models.

Although the calculation is often performed in a supercell eigenstate basis, it is effectively a real-space transport simulation because structural disorder is explicitly included in the atomic configuration. In this sense, real-space KG simulations directly connect atomistic structural models with quantum transport calculations.

To evaluate the dc conductivity, the treatment of the Dirac delta functions is a practical issue in KG calculations [26,43,44]. Because any finite supercell has a discrete energy spectrum, the delta functions must be replaced by broadened spectral functions associated with individual states [26,33,34] (see Fig. 4(a)). The form of the broadened spectral function is not unique; Lorentzian and Gaussian broadenings are commonly used depending on the numerical implementation. As a representative choice, a Lorentzian broadening can be written as

$$A_n(E) = \frac{1}{\pi}\frac{\eta}{(E-\varepsilon_n)^2+\eta^2},$$

where $A_n(E)$ is the broadened spectral function of eigenstate $n$, and $\eta$ is the spectral broadening parameter for this Lorentzian representation. Without such broadening, the energy-conservation condition imposed by the delta functions is rarely satisfied in a finite spectrum, so the calculated dc conductivity can vanish or become dominated by accidental level coincidences. The calculated dc conductivity can therefore depend sensitively on $\eta$. If $\eta$ is too small, the result is dominated by finite-size level discreteness. If $\eta$ is too large, distinct states are artificially mixed, and localization effects may be underestimated. Furthermore, $\eta$ should not be regarded solely as a numerical parameter. In predictive transport calculations, its relation to physical spectral broadening associated with thermal lattice fluctuations also requires careful consideration (see Fig. 4(b)).

The broadened spectral function approximates the delta-function energy-conservation condition in finite systems and enables the evaluation of the longitudinal dc conductivity, which can be expressed as [19,20,26]

$$\sigma_{\alpha\alpha}^{\mathrm{dc}} = \frac{2\pi e^2 \hbar}{\Omega} \int dE \left(-\frac{\partial f(E)}{\partial E}\right) \sum_{n,m} |\langle n|\hat{v}_\alpha|m\rangle|^2 A_n(E) A_m(E).$$

In this form, the dc conductivity is governed by the overlap of broadened spectral functions weighted by the corresponding velocity matrix elements. The spectral broadening therefore directly affects the conductivity obtained from finite-size KG calculations. In practice, evaluating this expression in finite disordered systems requires careful consideration of finite-size effects [43,44]. Near a mobility edge, the density of states, localization length, and velocity matrix elements can vary rapidly with energy, so the calculated conductivity can depend sensitively on the supercell size, spectral broadening, and sampling of disorder configurations. A systematic assessment of these factors is needed to distinguish intrinsic transport behavior from finite-size effects and numerical-smoothing-dependent contributions. For localized states, a finite $\eta$ relaxes the strict energy coincidence between discrete levels, allowing localized states to contribute to the dc conductivity in KG calculations. This contribution may resemble hopping-like transport at a phenomenological level, but it should not be identified directly with microscopic hopping transport. Phenomenological spectral broadening in conventional KG calculations does not, by itself, provide a complete description of phonon-assisted hopping conduction, which involves inelastic phonon-mediated transitions, including possible multiphonon processes, between localized states [8,9,29].

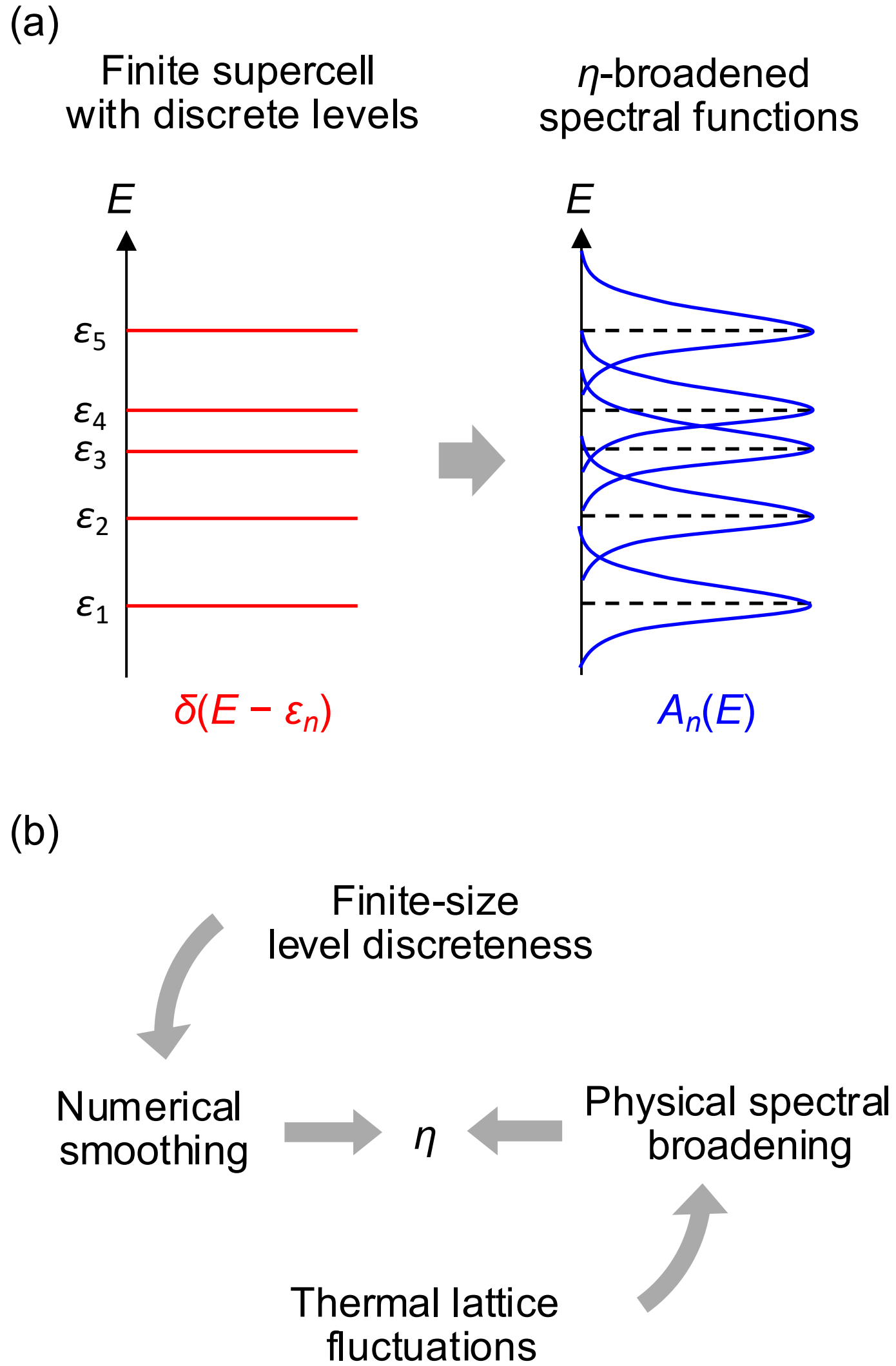

Fig. 4 Schematic illustration of the role of the broadening parameter $\eta$ in KG calculations for disordered materials. (a) In a finite supercell, the electronic spectrum is discrete, so the delta functions in the conductivity expression must be replaced by broadened spectral functions**.** (b) Numerical smoothing and physical spectral broadening have different origins, but their interplay requires careful interpretation in finite-cell KG calculations.

At finite temperature, the interpretation of finite-cell KG conductivity is further complicated by thermal lattice fluctuations. Molecular-dynamics-based Kubo–Greenwood (MD-KG) simulations incorporate thermal lattice fluctuations by evaluating the KG conductivity over an ensemble of molecular-dynamics snapshots [19,20,43,45]. In this procedure, an ensemble of atomic configurations is generated from molecular dynamics at a given temperature. For each snapshot, the Hamiltonian is constructed and the KG conductivity is evaluated. The temperature-dependent conductivity is then obtained by averaging over snapshots,

$$\sigma_{\alpha\alpha}(T) = \left\langle \sigma_{\alpha\alpha}^{(s)}(T) \right\rangle_s ,$$

where $\sigma_{\alpha\alpha}^{(s)}(T)$ is the conductivity calculated for molecular-dynamics snapshot $s$, and $\langle \cdots \rangle_s$ denotes averaging over the ensemble of snapshots. This averaging process naturally incorporates thermal fluctuations of bond lengths, bond angles, local coordination environments, and electronic couplings. As a result, MD-KG simulations can describe how finite-temperature lattice fluctuations modify both the energy spectrum and the velocity matrix elements. This snapshot-based treatment is especially useful when electron–phonon interactions can be approximated in the quasi-static limit [19,20]. In this limit, lattice vibrations are treated as slowly varying structural distortions relative to the electronic degrees of freedom. The effect of electron–phonon interactions is then incorporated through the thermal modulation of the Hamiltonian, rather than through explicit phonon-induced scattering rates between Bloch states. MD-KG therefore offers a practical approach for incorporating quasi-static electron–phonon effects in disordered systems. Its applicability, however, remains regime-dependent. In particular, snapshot averaging does not by itself determine the spectral broadening parameter $\eta$. Each finite-cell KG calculation still requires a finite $\eta$ to obtain a well-defined dc conductivity from the discrete spectrum, and the resulting conductivity can remain sensitive to its choice. The value of $\eta$ should therefore be tested carefully and interpreted consistently with the thermal lattice fluctuations already included through molecular-dynamics sampling.

Molecular-dynamics-based nonequilibrium Green's function (MD-NEGF) simulations provide an alternative atomistic method [46–48]. Similar to MD-KG, MD-NEGF directly incorporates realistic atomic structures and finite-temperature lattice fluctuations obtained from molecular dynamics. Unlike finite-cell KG calculations, MD-NEGF does not require replacing discrete energy-conservation delta functions by phenomenologically broadened spectral functions for numerical smoothing. Instead, transport is obtained from the transmission function of an open system, with the effects of thermal lattice fluctuations incorporated through molecular-dynamics snapshots as in MD-KG. However, extracting bulk transport properties often requires simulations over multiple device lengths and averaging over many disorder realizations, which can become computationally demanding for experimentally relevant amorphous systems. In contrast, MD-KG provides direct access to bulk conductivity within a linear-response framework and can be readily combined with large-scale atomistic models. The two methods are therefore complementary for atomistic transport simulations of amorphous materials: MD-NEGF provides a transmission-based description of transport in finite geometries, whereas MD-KG offers an efficient route for analyzing bulk conductivity, localization effects, and finite-temperature structural fluctuations.

## 5 Summary and outlook

This review has discussed electron transport in amorphous materials from the viewpoint of disorder and localization. In amorphous semiconductors, structural disorder modifies the spatial character of wavefunctions, producing band-tail states, localized states, and mobility edges that affect transport. The same localization-based picture can be used to understand transport in representative systems such as a-Si and AOSs. It also motivates atomistic transport methods, including real-space KG simulations, that can treat disorder and localization without relying on crystal momentum.

Despite these advances, predictive transport modeling in amorphous materials remains challenging. A key difficulty arises from the simultaneous presence of structural disorder and electron–phonon interactions, both of which can affect carrier localization and transport. Although recent MD-KG [19,20] and MD-NEGF [46,47] simulations provide practical routes for incorporating thermal lattice fluctuations through molecular-dynamics sampling, the physical connection between such fluctuations and the spectral broadening used in finite-cell KG calculations remains unclear. Another limitation is that these MD-based approaches often rely on a quasi-static treatment of phonons, in which thermal motion is represented as a sequence of static disordered configurations. Developing methods capable of incorporating dynamical electron–phonon interactions [43,45] beyond this approximation, as well as vertex corrections [33–35], remains a challenge for future research.

Another challenge is achieving both realistic structural modeling and reliable electronic structures in large-scale amorphous systems. Large-scale models are often required to capture the statistical nature of structural disorder, the spatial distribution of localized states, and medium-range structural correlations that govern carrier transport in amorphous materials. Recent advances in machine-learning interatomic potentials (MLIPs) have enabled the generation of larger amorphous structures and larger structural ensembles beyond the practical limits of conventional first-principles molecular dynamics [49,50]. Such large-scale structural modeling, however, must be accompanied by reliable electronic-structure descriptions. Key electronic properties relevant to amorphous-semiconductor transport, including band gaps, defect levels, and localized states, remain sensitive to the choice of electronic-structure method. Hybrid functionals [51] and the $GW$ approximation [52] can improve the prediction of these properties, but their computational cost often limits their application to large-scale transport simulations involving hundreds or thousands of atoms. Emerging methods beyond conventional semilocal DFT, including meta-GGA functionals such as r2SCAN [53,54], self-consistent DFT+$U$+$V$ [55,56], and machine-learning-based electronic-structure methods [57–60], may provide routes toward improved accuracy at an affordable computational cost.

Beyond the accuracy of the electronic structure itself, a related open issue is how the degree of structural disorder controls the crossover between extended and localized electronic states, and how this crossover is reflected in measurable transport properties. A more quantitative understanding of the relationships among disorder strength, spectral broadening, localization, and transport remains to be developed. Combining realistic structural models, reliable electronic structures, and advanced transport simulations will be important for developing predictive tools for amorphous-material design. Such developments should not only improve quantitative transport predictions but also provide a deeper physical understanding of how disorder and localization govern electron transport in amorphous materials.

**Acknowledgment**

The author thanks the members of the Materials and Device Theory Laboratory and the faculty members participating in the Solid State Physics Seminar for their valuable discussions and intellectual engagement, which motivated the preparation of this review.